# Satellite Motion in a Manev Potential with Drag

Samantha Kirk[1] Ioannis Haranas[2] Ioannis Gkigkitzis[1]

[1] *Departments of Mathematics, East Carolina University*
*124 Austin Building, East Fifth Street, Greenville*
*NC 27858-4353, USA*
*E-mail: gkigkitzisi@ecu.edu*

[2] *Department of Physics and Astronomy, York University*
*4700 Keele Street, Toronto, Ontario, M3J 1P3, Canada*
*E-mail:yiannis.haranas@gmail.com*

**Abstract**
In this paper, we consider a satellite orbiting in a Manev gravitational potential under the influence of an atmospheric drag force that varies with the square of velocity. Using an exponential atmosphere that varies with the orbital altitude of the satellite, we examine a circular orbit scenario. In particular, we derive expressions for the change in satellite radial distance as a function of the drag force parameters and obtain numerical results. The Manev potential is an alternative to the Newtonian potential that has a wide variety of applications, in astronomy, astrophysics, space dynamics, classical physics, mechanics, and even atomic physics.

## 1 Introduction

The Bulgarian physicist Professor Georgi Manev has introduced a classical potential in order to modify the celestial mechanics in accordance with the general-relativistic description: to describe the motion of a particle of mass $m$ in the static field of universal gravitation due to mass $M$, Manev (1924, 1925, 1930a, 1930b) replaced the mass $m$ with:

$$m = m_0 e^{\left(\frac{GM}{c^2 r}\right)} \tag{1}$$

where $m_0$ is an invariant and $G$ is the Newton's constant. This led to the following modification of the Newton's gravitational law takes the form (Hagihara 1975):

$$F_M = -\frac{GMm}{r^2}\left(1 + \frac{6GM}{c^2 r}\right) \tag{2}$$

where $m$ of a particle (satellite in our context), moving at the distance $r$ from a field-generating body (primary) of mass $M$. Therefore the corrected Newtonian potential becomes:

$$V = -\frac{GMm}{r}\left(1 + \frac{3GM}{c^2 r}\right) \tag{3}$$

where the Manev correction to the Newtonian potential is simply the term:

$$V_M = -\frac{3G^2 M^2 m}{c^2 r^2}. \tag{4}$$

as it's given in (Hagihara 1975; see also Maneff 1924, 1925, 1930a, b) Manev's potential was the candidate in the explanation of the historical anomaly of the Moon's perihelion precession (Maneff 1924, 1925, 1930a). Furthermore, one knows that the perihelion advance of the inner planets (especially that of Mercury) could not be entirely explained within the framework of the classical Newtonian law, even resorting to the theory of perturbations. The laws that exist usually answered the question that relates to the perihelion advance, but failed to explain other issues (such as the secular motion of the Moon's perigee). However, the modification of the Newtonian potential within a classical non-relativistic context is attributed to Manev (1930b). In the second decade of the 20th century, general relativity succeeded in explaining well such phenomena. In relation to general relativity potential corrections are first encountered in the Reissner - Nordström metric which now includes an extra correction to the potential of the form $V_{RN} \propto Q^2 / r^2$ (Nordström, 1918), where $Q$ is the charge measured in coulombs. Furthermore, the study of the anisotropic Manev problem might bring a contribution to a better understanding of the connection between quantum, relativistic, and classical mechanics (Craig et al. 1999, Diacu 2000; Diacu and Santoprete 2001). Ivanon and Pordanov (2005) present the modelling of general-relativistic results in a Newtonian framework that is now modified by a classical Manev type of potential. The authors analyze circular orbits around rotating and non-rotating black holes or primary massive bodies and finally they examine the parametrized post-Newtonian (PPN) formalism when they derive the potential that describes motion within the solar system. The Manev potential appearing in Eq. (3) is only a first order expansion. To detect the rotational effects of the mass the Manev potential will have to include more terms as in Eq. (35) in Ivanon and Pordanov (2005), otherwise, only the Schwarzschild geometry is recovered. Finally, in Adelberger et al. (2007) the authors investigate violations of the gravitational inverse-square law to constrain dilaton, radion, and chameleon exchange forces as well as arbitrary vector or scalar Yukawa interactions. In Chaliasos (2001) the author examines the pericenter precessions in a Reissner-Nordström metric. In Iorio (2012) the author investigates the orbital effects of a $r^{-2}$ in a Reissner - Nordström metric scenario. In particular he investigates natural as well as artificial satellites and solar system bodies perigee and perihelion motions, and also the periastron orbital motion for the orbiting compact object in Sag-A. Finally, in Calura et al. (1998) the authors work out the effects of interstellar dust drag in a Schwarzschild space–time using Gauss' planetary equation for the change of the semimajor axis w.r.t the true anomaly of the orbit. In Milani et al., (1987) the reader can find an extensive discussion on various non gravitational forces.

In this contribution we examine the indirect effect of the Manev potential onto the aerodynamic drag through its effects on the satellite's velocity that enters the drag formula. This is some type of mixed Manev-drag effect. For that we calculate the average energy dissipated by the satellite in one revolution, and from that the change in the orbital radial distance of the satellite is calculated for satellite of various aerodynamic coefficients, and orbital heights. Finally, we calculate the Manev-drag effect on the satellite's mean motion and period.

## 2 Satellite Motion and Drag

Satellites in low earth orbits have finite life times due to the effects of atmospheric drag. Although the atmosphere at the altitude of several hundred kilometers behaves like a perfect vacuum by earthly standards, the satellite must travel at very high speeds through low-density layers of this medium for years. The effects of drag are cumulative and finally become significant enough to take over the satellite's altitude reduction. For a satellite orbiting at a low altitude the acceleration due to drag is given by (Vallado, 2007):

$$a_d = -\frac{1}{2}\frac{C_d A_s}{m_s}\rho(r)v_{S_{rel}}^2 \qquad (3)$$

where $C_d$ is the coefficient of drag , $A_s$ is the presented area of the satellite, and $m_s$ and $\vec{v}_{S_{rel}} = \vec{v}_s - (\vec{\omega}_A \times \vec{r})$ are the satellite's mass and relative velocity with respect to a rotating atmosphere respectively, and $\rho(r)$ is the atmospheric density function, and $\vec{\omega}_A$ is the angular rotation velocity vector of the atmosphere and $\vec{r}$ radial vector of the satellite. In this paper the rotational velocity of the Earth's atmosphere is neglected. The first result of the drag acting on the satellite's orbit is to circularize the orbit. This takes place at perigee where the effect of the drag is more evident, since the satellite is at its deepest stage inside the atmospheric layers. At this point, the satellite slows down and lowers its apogee height until it becomes nearly equal to the perigee height. To proceed with our calculations we will assume that the orbit of the satellite has been circularized and thus study the behavior of its semimajor axis. If the orbit is circular, the air drag acting on the satellite is in the tangential direction. For the atmospheric density function, we will assume the exponential function given below (Vallado, 2007):

$$\rho(h) = \rho_0 e^{-\frac{(h_{el}-h_0)}{H}} \qquad (4)$$

where, a reference density $\rho_0$ is used with a reference altitude, $h_0$ is the actual altitude above the ellipsoid $h_{el}$ and $H$ is the atmospheric scale height. Many authors have investigated the non-Newtonian effects on satellites, planets. For example a good and comprehensive study of the Lens-Thirring effect in the solar system the reader can referrer themselves to Iorio et al., (2011). Similarly, in Iorio (2002, 2010) the author investigates the possibility of constraining the hypothesis of a fifth force at the length scale of two Earth's radii by investigating the effects of a Yukawa gravitational potential on the orbits of the laser-ranged LAGEOS and LAGEOS II satellites respectively, as well as he studies the effects of neutral and charged drag on the inclination of the LARES satellite and with respect to LAGEOS and LAGEOS II. Finally, in Haranas et al., (2011) they investigate the effects of the non-Newtonian Yukawa potential on the anomalistic time of the celestial bodies, such as Mercury, the pulsar PSR 1913+16, and the Earth orbiting satellite GRACE - A. For other performed or proposed studies of non-Newtonian gravity with GRACE, see Larson et al. (2007), and also Iorio (2012).

### 3 Circular Motion and the Calculation of the Radial Distance Change

Assuming circular motion we can calculate the circular orbital velocity $r = a$ and $e = 0$ to be:

$$v_{tot_c}^2 = v_N^2 + v_M^2 = \frac{GM}{r} + \frac{6G^2M^2}{c^2r^2},\tag{5}$$

where $v_N$ and $v_M$ are the Newtonian and the Manev components of the total orbital circular velocity, and $M$ is the mass of the primary body. Air drag changes the energy of the orbit, since the drag acceleration $a_d$ does work on the satellite at $a_d.V = -v\ a_d$ where the negative sign is due to the fact that drag opposes the velocity. Next we obtain the total energy per unit mass using the relation:

$$\mathcal{E}_{tot} = \frac{1}{2}v_{tot}^2 - V_{tot}.\tag{6}$$

Eliminating $v_{tot}^2$ using Eq. (5) and also using Eq. (3) we obtain:

$$\mathcal{E}_{tot} = -\frac{GM}{2r}\tag{7}$$

To proceed with the calculation of the change in radial distance, let us calculate the change of energy in one revolution. Integrating around for a full revolution we write that:

$$\Delta\mathcal{E} = -\int_0^{2\pi}\frac{A_s C_d \rho_0}{2}e^{-\frac{(h_d - h_0)}{H}}v_{tot_c}^2 r\mathrm{d}\theta.\tag{8}$$

Using Eq. (5) we eliminate $v_{tot_c}^2$ in Eq. (8) and therefore have that:

$$\Delta \mathcal{E} = -\int_0^{2\pi} \frac{A_s C_d \rho_0}{2} e^{-\frac{(h_d - h_0)}{H}} \left( \frac{GM}{r} + \frac{6GM}{c^2 r^2} \right) r \mathrm{d}\theta \,, \tag{9}$$

furthermore, assuming a small radial change per revolution, and also a constant density over one satellite period, the energy loss becomes:

$$\Delta \mathcal{E} = -\pi \, GM C_d A_s \rho_0 e^{-\frac{(h_d - h_0)}{H}} \left( 1 + \frac{GM}{c^2 r} \right). \tag{10}$$

Next, writing the total energy of the satellite we obtain:

$$\Delta \mathcal{E} = m_s \Delta \left( -\frac{GM}{2r} \right) = \frac{GM m_s}{2r^2} \Delta r \,, \tag{11}$$

and by solving for $\Delta r$ we find the total change in the satellite radial distance to be:

$$\Delta r = -\frac{2\pi \, r^2}{m_s} C_d A_s \rho_0 e^{-\frac{(h_d - h_0)}{H}} \left( 1 + \frac{GM}{c^2 r} \right), \tag{12}$$

and in particular the Manev contribution becomes:

$$\Delta r = -2\pi \left( \frac{GM C_d A_s}{m_s c^2} \right) \rho_0 \; r e^{-\frac{(h_d - h_0)}{H}} \,, \tag{13}$$

where $\Delta r$ is the change in radial distance due to the Newtonian potential only, and therefore we have:

$$\Delta r_N = -\frac{2\pi C_d A_s}{m_s} r^2 \rho_0 e^{-\frac{(h_d - h_0)}{H}} \,. \tag{14}$$

Using equations (14), (15), and (16) we obtain that:

$$\frac{\Delta r_{tot}}{\Delta r_M} = -\left( 1 + \frac{c^2 r}{GM_E} \right) = -\left( 1 + \frac{2r}{R_{gr_E}} \right) = -\left( 1 + \frac{c^2}{v_{cir}^2} \right), \tag{15}$$

and

$$\frac{\Delta r_{tot}}{\Delta r_N} = -\left( 1 + \frac{GM_E}{c^2 r} \right) = -\left( 1 + \frac{R_{gr_E}}{2r} \right)^{-1} = -\left( 1 + \frac{v_{cir}^2}{c^2} \right), \tag{16}$$

from which we also obtain the relation

$$\frac{\Delta r_N}{\Delta r_M} = \left(1 + \frac{c^2 r}{GM_E}\right)\left(1 + \frac{GM_E}{c^2 r}\right)^{-1} = \left(1 + \frac{2r}{R_{g_{r_E}}}\right)\left(1 + \frac{R_{g_{r_E}}}{2r}\right)^{-1} = \left(1 + \frac{c^2}{v_{cir}^2}\right)\left(1 + \frac{v_{cir}^2}{c^2}\right),$$ (17)

where $R_{g_{r_E}} = \frac{2GM}{c^2}$ is the gravitational radius of the Earth. Similarly, we obtain that:

$$\frac{\Delta E_{tot}}{\Delta E_M} = -\left(1 + \frac{c^2 r}{GM_E}\right) = -\left(1 + \frac{2r}{R_{g_{r_E}}}\right) = -\left(1 + \frac{c^2}{v_{cir}^2}\right),$$ (18)

$$\frac{\Delta E_{tot}}{\Delta E_N} = -\left(1 + \frac{GM_E}{c^2 r}\right) = -\left(1 + \frac{R_{g_{r_E}}}{2r}\right) = -\left(1 + \frac{v_{cir}^2}{c^2}\right),$$ (19)

$$\frac{\Delta E_N}{\Delta E_M} = \left(1 + \frac{c^2 r}{GM_E}\right)\left(1 + \frac{GM_E}{c^2 r}\right)^{-1} = \left(1 + \frac{2r}{R_{g_{r_E}}}\right)\left(1 + \frac{R_{g_{r_E}}}{2r}\right)^{-1} = \left(1 + \frac{c^2}{v_{cir}^2}\right)\left(1 + \frac{v_{cir}^2}{c^2}\right).$$ (20)

Next let us examine the change of the mean motion of the satellite $n$ using the equation for the mean motion $n^2 = GM/r^3$ we obtain that:

$$2n\Delta n = -\frac{3GM_E}{r^4}\Delta r \qquad .$$ (21)

Substituting Eq. (14) in Eq. (22) we obtain that that the total change in the satellite mean motion becomes:

$$\Delta n_{tot} = \frac{3\pi}{n}\left(\frac{GM}{r^2 m_s}\right)C_d A_s \rho_0 e^{-\frac{(h_d - h_0)}{H}}\left(1 + \frac{GM}{c^2 r}\right)$$ (22)

The Newtonian contribution to the mean motion change is:

$$\Delta n_N = \frac{3\pi}{n}\left(\frac{GM}{r^2 m_s}\right)C_d A_s \rho_0 e^{-\frac{(h_d - h_0)}{H}},$$ (23)

and the Manev contribution becomes:

$$\Delta n_M = \frac{3\pi}{n}\left(\frac{G^2 M^2}{c^2 r^3 m_s}\right)C_d A_s \rho_0 e^{-\frac{(h_d - h_0)}{H}}.$$ (24)

Furthermore, for the change in the orbital period we use $P^2 = 4\pi^2 a^{3/2}/GM_E$, and therefore:

$$\Delta P = \frac{6\pi^2}{GM_E}\frac{r^2}{P}\Delta r.$$ (25)

Substituting $\Delta r$ from Eq. (14) into Eq. (15):

$$\Delta P_{tot} = -\frac{12\pi^3}{P} \frac{r^4}{GM_E m_s} AC\rho_0 e^{\frac{(h_{el}-h_0)}{H}}\left(1 + \frac{GM}{c^2 r}\right),$$  (26)

and therefore the Newtonian and Manev parts respectively become

$$\Delta P_N = \frac{12\pi^3}{P} \frac{r^4}{GM_E m_s} AC\rho_0 e^{\frac{(h_{el}-h_0)}{H}},$$  (27)

$$\Delta P_M = \frac{12\pi^3}{P} \frac{r^3}{m_s c^2} AC\rho_0 e^{\frac{(h_{el}-h_0)}{H}}.$$  (28)

From Eqs. (23)-(25) we obtain that:

$$\frac{\Delta n_{tot}}{\Delta n_N} = \left(1 + \frac{GM_E}{c^2 r}\right) = \left(1 + \frac{R_{g_{r_E}}}{2r}\right) = \left(1 + \frac{v_{cir}^2}{c^2}\right),$$  (29)

similarly

$$\frac{\Delta n_{tot}}{\Delta n_M} = \frac{c^2 r}{GM_E}\left(1 + \frac{GM_E}{c^2 r}\right) = \frac{2r}{R_{g_{r_E}}}\left(1 + \frac{R_{g_{r_E}}}{2r}\right) = \frac{c^2}{v_{cir}^2}\left(1 + \frac{v_{cir}^2}{c^2}\right).$$  (30)

and

$$\frac{\Delta n_N}{\Delta n_M} = \frac{c^2 r}{GM_E} = \frac{r}{2R_{g_{r_E}}} = \frac{c^2}{v_{cir}^2}$$  (31)

Using Eqs.(28)-(30) we obtain:

$$\frac{\Delta P_{tot}}{\Delta P_N} = \left(1 + \frac{GM_E}{c^2 r}\right) = \left(1 + \frac{R_{g_{r_E}}}{2r}\right) = \left(1 + \frac{v_{cir}^2}{c^2}\right),$$  (32)

and also we have:

$$\frac{\Delta P_{tot}}{\Delta P_M} = \frac{c^2 r}{GM_E}\left(1 + \frac{GM_E}{c^2 r}\right) = \frac{2r}{R_{g_{r_E}}}\left(1 + \frac{R_{g_{r_E}}}{2r}\right) = \frac{c^2}{v_{cir}^2}\left(1 + \frac{v_{cir}^2}{c^2}\right),$$  (33)

and finally:

$$\frac{\Delta P_N}{\Delta P_M} = \frac{c^2 r}{GM_E} = \frac{r}{2R_{g_{r_E}}} = \frac{c^2}{v_{cir}^2} .$$  (34)

In a recent work by Kezerashvili and Poritz, (2010) the authors have calculated the correction $P_Q$ to the Keplerian orbital period $P$ resulting from a Reissner-Nordström metric. This is a study of general relativistic effects on bound orbits of solar sails. Similarly in Iorio and Lichtenegger (2005) the authors examine the possibility of singling out the gravitomagnetic effect on the mean longitudes by means of a suitable space-based environment. They deal with various competing effects which could introduce errors and bias.

## 4 Numerical Results and Discussion

At this point we want to proceed with some numerical calculations. This also emanates from several current theories predict $r^{-2}$ corrections to the potential. As an example in Adelberger et al. (2007) the author investigate violations of the gravitational inverse-square law to constrain dilaton, radion, and chameleon exchange forces as well as arbitrary vector or scalar Yukawa interactions. In order to quantitatively evaluate our findings let us now consider a satellite in a circular orbit i.e. $e = 0$ at a radial distance from the center of the Earth where $r = 7125$ km. The orbital altitude of the satellite is equal to $h = r - R_E = 7125 - 6378 = 747$ km, and therefore the density becomes:

$$\rho = 3.614 \times 10^{-14} e^{\frac{(747-700)}{88.667}} \cong 2.123 \times 10^{-14} \, \text{kg/m}^3, \tag{35}$$

where the nominal density has been taken to be $\rho_0 = 3.614 \times 10^{-14} \, \text{kg/m}^3$ (Vallado, 2007) and the scale height $H = 88.667$ km have been used (Vallado, 2007). We take satellites of masses $m_s = 900$ kg, aerodynamic drag coefficients $C_d = 2.0, 2.1, 2.2, 2.3, 2.5$ respectively, and $M = M_E = 5.94 \times 10^{24}$ kg. Our results are tabulated in table 1 below.

**Table 1** Change of the radial distance of a satellite moving in a Manev gravitational field in a Circular orbit around the Earth

| Satellite Mass $m$ [kg] | Satellite Radial Distance $r$ [km] | Aerodynamic Drag Coefficient $C_D$ | $\Delta r_N$ [m/rev] | $\Delta r_M$ [m/rev] | $\Delta r_{tot}$ [m/rev] |
|---|---|---|---|---|---|
| 900 | 7125.3489 | 2.0 | -0.0004513 | $-2.788 \times 10^{-12}$ | -0.0004513 |
| | | 2.1 | -0.0004740 | $-2.930 \times 10^{-12}$ | -0.0004740 |
| | | 2.2 | -0.0004964 | $-3.067 \times 10^{-12}$ | -0.0004964 |
| | | 2.3 | -0.0005190 | $-3.206 \times 10^{-12}$ | -0.0005190 |
| | | 2.4 | -0.0005415 | $-3.346 \times 10^{-12}$ | -0.0005415 |

**Table 2** Change of the radial distance of a satellite moving in a Manev gravitational field in a circular orbit around the Earth.

| Satellite Mass $m$ [kg] | Satellite Radial Distance $r$ [km] | Aerodynamic Drag Coefficient $C_D$ | $\Delta r_N$ [m/rev] | $\Delta r_M$ [m/rev] | $\Delta r_{tot}$ [m/rev] |
|---|---|---|---|---|---|
| 900 | 6728.137 | 2.0 | -18.047 | $-1.180\times10^{-8}$ | -18.047 |
| | | 2.1 | -18.950 | $-1.240\times10^{-8}$ | -18.950 |
| | | 2.2 | -19.852 | $-1.299\times10^{-8}$ | -19.852 |
| | | 2.3 | -20.755 | $-1.358\times10^{-8}$ | -20.755 |
| | | 2.4 | -20.660 | $-1.417\times10^{-8}$ | -20.660 |

**Table 3** Change of the mean motion of a satellite moving in a Manev gravitational field in a circular orbit around the Earth.

| Satellite Mass $m$ [kg] | Satellite Radial Distance $r$ [km] | Aerodynamic Drag Coefficient $C_D$ | $\Delta n_N$ [rad/s] | $\Delta n_M$ [rad/s] | $\Delta n_{tot}$ [rad/s] |
|---|---|---|---|---|---|
| 900 | 6728.137 | 2.0 | $4.590\times10^{-9}$ | $3.002\times10^{-18}$ | $4.590\times10^{-9}$ |
| | | 2.1 | $4.818\times10^{-9}$ | $3.153\times10^{-18}$ | $4.818\times10^{-9}$ |
| | | 2.2 | $5.050\times10^{-9}$ | $3.302\times10^{-18}$ | $5.050\times10^{-9}$ |
| | | 2.3 | $5.277\times10^{-9}$ | $3.453\times10^{-18}$ | $5.277\times10^{-9}$ |
| | | 2.4 | $5.070\times10^{-9}$ | $3.603\times10^{-18}$ | $5.070\times10^{-9}$ |

**Table 4** Change of the period of a satellite moving in a Manev gravitational field in a circular orbit around the Earth.

| Satellite Mass $m$ [kg] | Satellite Radial Distance $r$ [km] | Aerodynamic Drag Coefficient $C_D$ | $\Delta P_N$ [s/rev] | $\Delta P_M$ [s/rev] | $\Delta P_{tot}$ [s/rev] |
|---|---|---|---|---|---|
| 900 | 6728.137 | 2.0 | -0.0222 | $-1.451\times10^{-11}$ | -0.0222 |
| | | 2.1 | -0.0233 | $-1.523\times10^{-11}$ | -0.0233 |
| | | 2.2 | -0.0244 | $-1.596\times10^{-11}$ | -0.0244 |
| | | 2.3 | -0.0255 | $-1.670\times10^{-11}$ | -0.0255 |
| | | 2.4 | -0.0266 | $-1.741\times10^{-11}$ | -0.0266 |

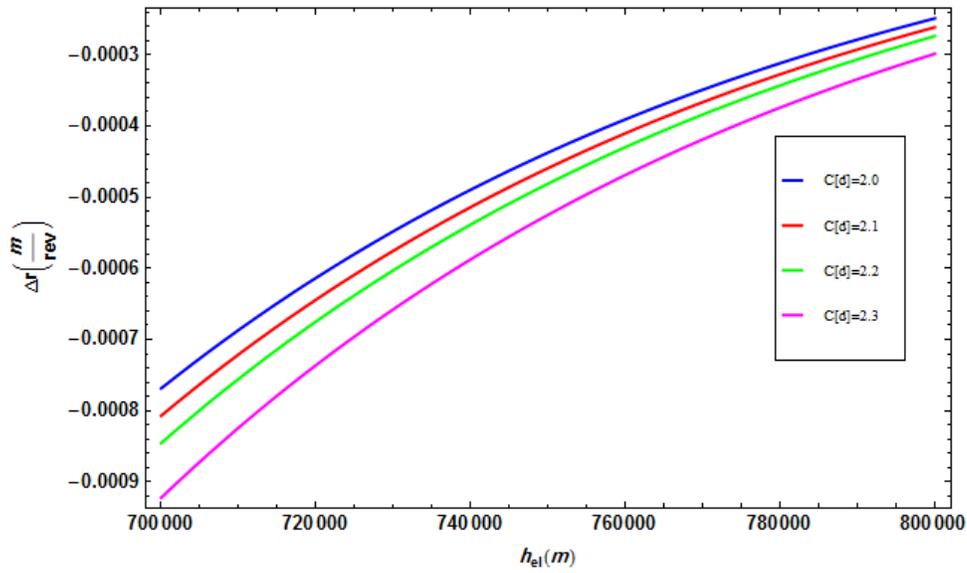

**Fig. 1** Radial distance change $\Delta r$ as a function of orbital reference altitude $h_{el}$ for a satellite of a surface area $A_s = 3$ m$^2$ and mass $m_s = 900$ kg in a circular orbit.

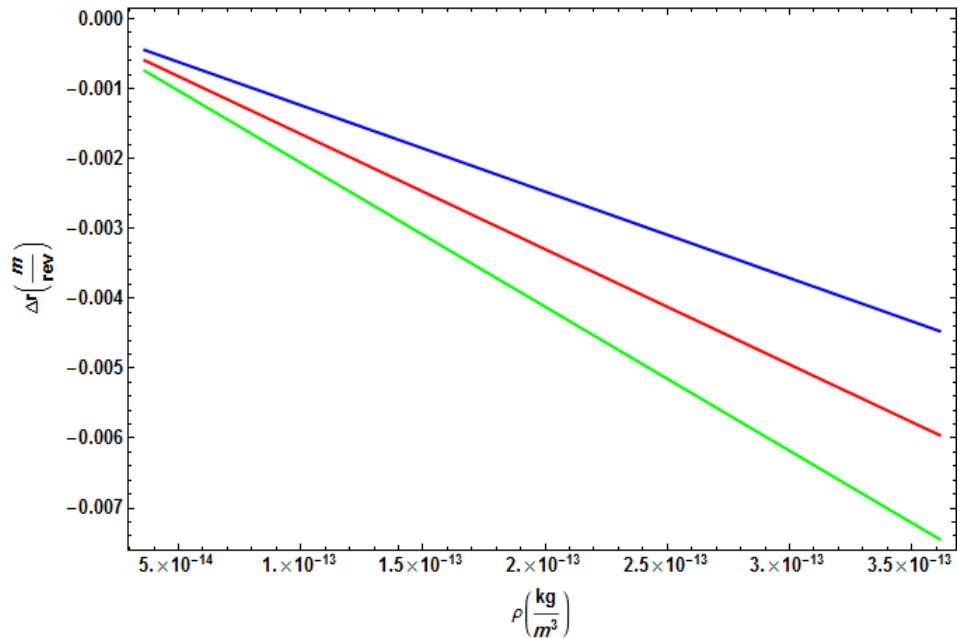

**Fig. 2** Radial distance change $\Delta r$ as a function of atmospheric density in the reference altitude $h = 700$-$800$ km, for $m = 1000$ kg satellite of various surface areas in circular orbit. Blue = 3m$^2$, red = 4m$^2$, green = 5m$^2$.

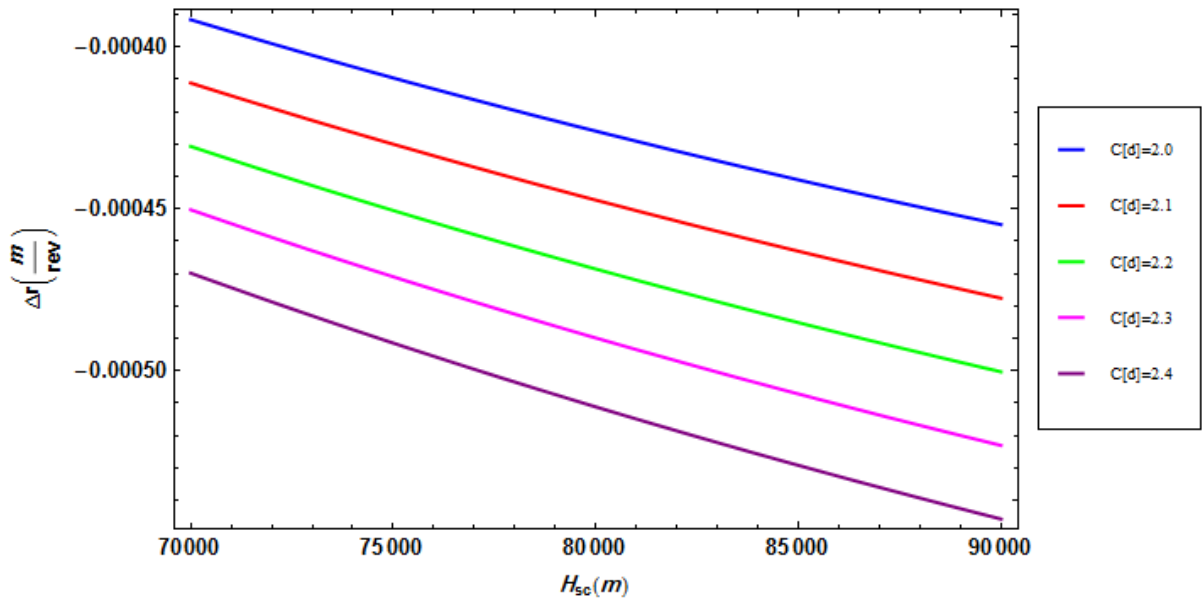

**Fig. 3** Radial distance $\Delta r$ as a function of atmospheric density in the reference altitude $h$= 700-800 km, for a $m$ = 1000 kg satellite in circular orbit.

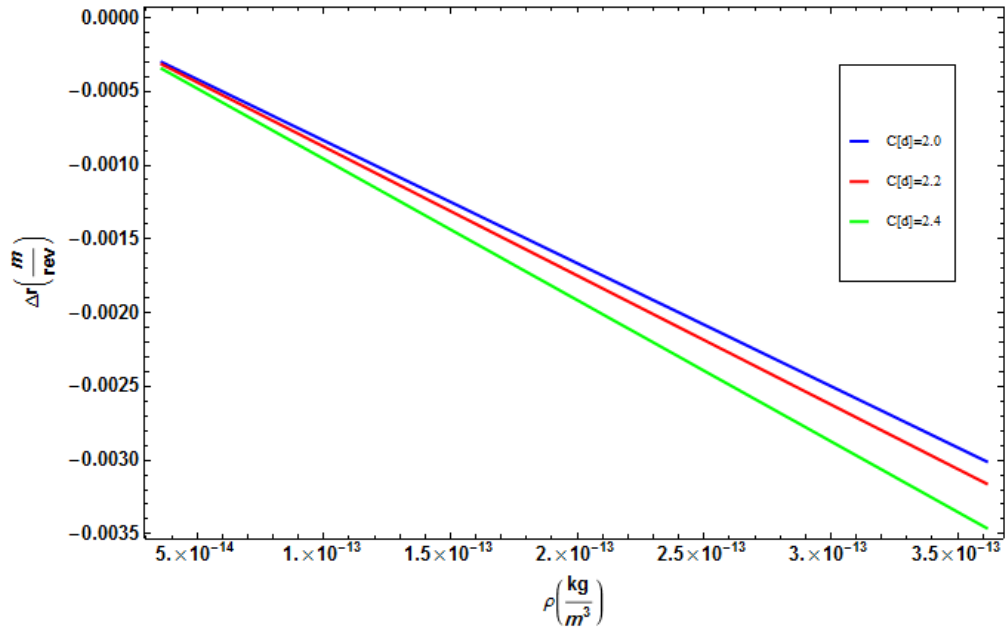

**Fig. 4** Radial distance $\Delta r$ as a function of atmospheric density in the reference altitude $h$= 700-800 km, for a $m$ = 900 kg satellite in circular orbit, for various aerodynamic drag coefficients.

We have derived the change in radial distance of a satellite due to drag and a Manev type of gravitational field. We have assumed a circular orbit satellite at the orbital altitude $h = 747.2119$ km. In a circular orbit at this altitude the Manev field for 900 kg satellite introduces a correction to the drag force that is equal to

$$F_{d_M} = \frac{1}{2} C_d A_s \rho v_{M_C}^2 = \frac{3 G^2 M_E^2}{c^2 r^2} C_d A_s \rho_0 e^{-\frac{(h_d - h)}{H}} \cong O\left(10^{-12}\right) \text{ N} = 1 \text{pN} \tag{36}$$

Or equivalently an acceleration that is equal to:

$$a_{d_M} = \frac{1}{2 m_s} C_d A_s \rho v_{M_C}^2 = \frac{3 G^2 M_E^2}{c^2 r^2 m_s} C_d A_s \rho_0 e^{-\frac{(h_d - h)}{H}} \cong O\left(10^{-15}\right) \text{ m s}^{-2} = 1 \text{fN} \tag{37}$$

This acceleration is equivalent to $1.0 \times 10^{-7}$ µgal which is about seven orders of magnitude less that the detectability limit of today's satellite accelerometers and will be hard to detect. Li et al. (2007) have found that the performance on-orbit of the drag-free translation control system satisfies the requirements of the GP-B science experiment. The authors have found a $40 \times 10^{-12}$ m/s$^2$ acceleration along the roll axis of a drag free satellite. Maybe future technological improvements will be able to achieve this limit. Therefore, we see that in the case of the orbiting satellite under consideration, the most important contribution to the change of the radial distance per revolution comes from the effect of the Newtonian part of the potential in the circular velocity of the orbiting satellite. The contribution of the Manev potential in the change of the radial distance significantly increases in lower orbit satellites. This is due to the fact that the atmospheric density and radial distance increase and decrease significantly. At lower radial distance the satellite will also have a higher total circular orbital velocity, and therefore the drag will be higher. For an orbit at $h = 350$ km results to an atmospheric density $\rho = 9.518 \times 10^{-12}$ kg/m$^3$ and the corresponding Manev force and acceleration contribution become:

$$F_{d_M} = \frac{1}{2} C_d A_s \rho v_{M_C}^2 = \frac{3 G^2 M_E^2}{c^2 r^2} C_d A_s \rho_0 e^{-\frac{(h_d - h)}{H}} \cong O\left(10^{-8}\right) = 10 \text{ nN} \tag{38}$$

$$a_{d_M} = \frac{1}{2 m_s} C_d A_s \rho v_{M_C}^2 = \frac{3 G^2 M_E^2}{c^2 r^2 m_s} C_d A_s \rho_0 e^{-\frac{(h_d - h)}{H}} \cong O\left(10^{-6}\right) = 1 \text{µgal} \tag{39}$$

The Manev contribution will become significantly higher if the mass of the primary body increases. Just to get an idea let us assume that it is possible to have a satellite orbiting in a circular orbit at $r = 1.7 \times 10^5$ km, a body that is possible to have the same atmosphere with the Earth's atmosphere, and mass equal to that of the sun. In this case the acceleration due to Manev gravitational field contribution becomes:

$$a_{d_M} = O\left(10^{-8}\right) = 1\,\mu\text{gal}. \qquad (40)$$

With the help of future and more sensitive technology, more massive primary bodies might help test and detect the Manev effects. Observation of the Manev effect in circular orbit satellite will require $10^8$ revolutions or 17460 years of observing, at the end of which the effect will exhibits a radial change $\Delta r = -1.417$ m. This practically impossible, since such a satellite life time does not exist. Similarly, it will take 3170 satellite revolutions in order to observe a satellite period reduction of -1.714 s. In relation to the mean motion it will take approximately 8 times the age of the universe in order to observe a change in mean motion equal to 3.603 rad.

## 5 Conclusion

Several modern theories of today predict additional $r^{-2}$ corrections to the Newtonian potential, thus resulting in the gravitational inverse-square law constraining dilaton, radion, and chameleon exchange forces as well as arbitrary vector or scalar Yukawa interactions. In this paper, we have calculated the drag effect exerted on a satellite in a circular orbit when a Manev $r^{-2}$ correction to the Newtonian potential is taken into account. The effect of the Manev part enters through the correction to the satellite's velocity from which the drag force depends. For a satellite with mass $m = 900$ kg and aerodynamic drag coefficient $C_d = 2.4$ orbiting at 7125 km we obtained $\Delta r_M = -3.346 \times 10^{-12}$ m/rev, which is $6.0 \times 10^{-9}$ times smaller than the Newtonian counterpart. Low orbit satellites demonstrate a higher Manev drag effect which for a 900 kg satellite of the same aerodynamic drag coefficient reaches up $-14.17$ nm/rev. More massive bodies such as planets or stars can be used to examine the Manev effect in satellite motion with drag.

**Acknowledgements** The authors would like to thank the anonymous reviewer for his valuable comments and suggestions that helped to improve this manuscript considerably.